\documentclass{article}
\DeclareMathAlphabet{\mathsfsl}{OT1}{cmss}{m}{sl}
\def\beq{\begin{equation}}
\def\eeq{\end{equation}}
\def\cm1{$\rm cm^{-1}$}
\def\DE{D\kern-0.75em \raisebox{1.0pt}{=}\ }

\newcommand{\hangpar}{\noindent\hangindent.5in}%
\begin{document}
\def\Rh{\rule{20.0pt}{0.0pt}}
\def\Rv{\rule[-0.1in]{0.0pt}{20.0pt}}
\def\SC{{\mathsfsl{C}}}
\def\SQ{{\mathsfsl{Q}}}
\begin{center}
{\bf CALCULATIONS AT SERIES LIMITS IN ONE-ELECTRON SYSTEMS}

\vspace{0.2 in}
{\it Charles R. Cowley}

{\it Department of Astronomy, University of Michigan}
\end{center}

\vspace{0.2 in}
{\it Introduction}
\vspace{0.2 in}

The behavior of atoms in high quantum levels is generally well
understood---in principle.  For sufficiently large values of the
principle quantum number, discrete levels that would exist for an
isolated atom,  do not. They are said to be dissolved, or quenched by
perturbations from surrounding atoms and ions. In practice, the
calculation of relevant atomic properties can become quite involved,
especially if one strives for accuracy.  Detailed references may be
found in the papers by Stehl\'{e} and Jacquemot$^1$, 
Hummer and Mihalas$^2$, or the book by Griem$^3$.

We draw attention here to a general principle that should be
of practical value in the calculation of opacity near the 
series limits.  
We show that as soon as the Lyman or Balmer lines overlap and the
individual profiles wash out, the net absorption cross
section {\it rapidly} approaches the value at the series limit.
It has been known for more than seventy years that the line and
continuous opacities merge smoothly.  We are unaware of a discussion
of the {\it rapidity} of this convergence, which has a practical
value for the calculation of opacity.  

Independent of the difficult question of the relative
contributions of line and continuous opacity, we can 
get a good estimate of their sum.  The accuracies of estimates for
two  series are discussed below. 
We offer the following thoughts as a computational aid.
They may also be of heuristic value.

\vspace{0.2 in}
{\it A Mean Line Absorption Cross Section}   
\vspace{0.2 in}

Let us begin by considering the line opacity in a series.
We take the Lyman series for simplicity.  For lines with
wavenumbers greater than that of L$_\alpha$, we may define
a mean cross section for the line L$_n$ in the following
way.  Associate an interval, 
$\Delta\tilde{\nu}_n$ cm$^{-1}$, with 
each line that extends halfway to the neighbors on either
side.  With the Rydberg formula, we have 

\begin{equation}
\Delta\tilde{\nu}_n = {1\over 
2}[\tilde{\nu}_{n+1}+\tilde{\nu}_n] - {1\over 2}
[\tilde{\nu}_n + \tilde{\nu}_{n-1}] 
= \left[2n\cdot Ry\over{(n-1)^2\cdot (n+1)^2}\right ].
\end{equation}                      

Clearly, $\Delta\tilde{\nu}_n$ represents the interval 
most logically associated with an individual line in the 
series.  We can use it to define a mean 
line photoabsorption 
cross section $<\sigma_n>$, such that

\begin{equation}          
<\sigma_n> \cdot \Delta\tilde{\nu}_n = 
{\pi e^2\over \mu c^2}f_{mn}.
\end{equation}

\noindent The symbols $e$ and $c$ have their usual meanings.  
Here, $\mu = m_e/(1+m_e/m_p)$ is the reduced mass, which we
must use to obtain the desired accuracy below.  Values of the
physical constants were taken from the NIST website$^4$.  We
use $e({\rm esu}) = e({\rm SI})*c/10$,  $a_H = a_0/(1+m_e/m_p)$,
and $Ry = e^2/(2a_Hhc)$.  

There 
would be a factor of $\lambda^2$ on the right of Equation (2) had we used
wavelengths rather than wavenumbers to define the interval
associated with the $n^{th}$ series member.  

Oscillator strengths for the Lyman lines have a simple closed
form (cf. Bethe and Salpeter$^5$ p. 263):
\begin{equation}
f_{1s,np} = {{2^8n^5(n-1)^{2n-4}}\over {3(n+1)^{2n+4}}}, 
\end{equation} 
which was used to derive the entries in column 2, of Table 1, 
which extend somewhat beyond the tabulation of Green, Rush,
and Chandler$^6$.  The last two figures given in this 
reference and in our Table 1 are of heuristic and illustrative
value only, since relativistic effects are of the order of
$(1/137)^2$ and apply to the real atom.  We have used non-relativistic
formulae throughout.  The last entry gives the
continuous absorption cross section calculated at the 
Lyman series limit.  The corresponding gaunt factor is  
0.797301.
Karzas and Latter$^7$ gave 0.7973.

\begin{table}
\caption{Mean Cross Sections (Power of ten shown in parentheses)}
\begin{center}
\begin{tabular}{c| l l| l l}\hline
        &\multicolumn{2}{c|}{Lyman}&\multicolumn{2}{c}{Balmer}  \\
n(upper)& f-value  & $<\sigma_n>$ cm$^2$& f-value  & $<\sigma_n>$ cm$^2$ \\ \hline \hline
 10     & 1.60537(-3)&   6.35353(-18)&3.85061(-3)&1.52395(-17)\\
 15     & 4.68624(-3)&   6.32994(-18)&1.07022(-3)&1.44560(-17)\\
 20     & 1.96675(-4)&   6.32172(-18)&4.41644(-4)&1.41958(-17)\\
 25     & 1.00456(-4)&   6.31796(-18)&2.23836(-4)&1.40776(-17)\\
 30     & 5.80583(-4)&   6.31587(-18)&1.28882(-4)&1.40141(-17)\\
 60     & 7.24117(-6)&   6.31236(-18)&1.59530(-5)&1.39067(-17)\\
 $\infty$&  &            6.31119(-18)&  &1.38712(-17)  \\ \hline

\end{tabular}            

\end{center}
\end{table}

Even for L$_{10}$, $<\sigma_{10}>$ differs from the cross 
section at the series limit by less than one per cent.  
Calculations for the Balmer series show a similar
convergence to the series limit, 
$1.387\cdot 10^{-17}$ cm$^2$
though less rapid.  The error at H$_{10}$ is nearly 10\%.
This is entirely consistent with the beautiful experimental 
work of Wiese and coworkers$^8$ which shows a drop in the 
intensity of about 10\% between $\lambda$3700 (essentially,
H$_{10}$) and the Balmer limit.

By H$_{20}$, the difference is only 2\%.  This line
($\lambda_{\rm air}$3682.81) is still 36.83\AA\ from the
series limit, $\lambda_{\rm air}3645.98$, and anyone
who analyzes lines in this region needs to have the 
H I opacity accurately calculated.
For conditions in the mean solar photosphere,
the Balmer lines overlap and flatten near the position of  
H$_{20}$, so 
the total opacity should be within 2\% of that at the series
limit from this point to the Balmer head. 

Calculations of the opacity near the 
Balmer and Lyman limits are therefore not sensitive to the 
difficult questions involving the ``existence'' of 
high upper levels.  Whatever the relative proportions of
line and continuous opacity, the sum must approach the
value at the head of the series. 

The rigorous approach to calculations in this region is 
described for example, by Stehl\'{e} and  Jacquemot$^1$
and in elegant detail, by Seaton$^9$, who also treats
the non-hydrogenic case.  At each 
wavelength, some fraction of the opacity is calculated as
line opacity, and the complement is calculated as 
continuous absorption.  The recipe for the division 
follows from the probability that a given level $n$ will
exist.  This probability is given in the references 
already cited,
and can be somewhat tedious.  It is therefore well for 
anyone coding this problem anew to be aware that the total 
opacity must merge rapidly to that at the series limit.
We have shown that this must be so for the Lyman and Balmer
series.  Higher series should be investigated.  If the 
convergence to the series limit is not rapid,
the ideas outlined here will require modification
to be useful computationally.
\vspace{0.2 in}

{\it A Proof}

\vspace{0.2 in}

We sketch a proof of the result that 
the quotient $f_{mn}/\Delta\tilde{\nu}_n$ reaches a limit 
for large $n$ that is independent of $n$, and which 
converges to the continuous absorption cross section at 
the series limit.  It is only necessary to divide the f-value
given by Equation (3) by $\Delta\tilde{\nu}$, which appears
on the right of Equation (1), and take the limit as $n$
approaches infinity.  We can then readily show that 

\begin{equation}
\sigma_{\rm lim} = \lim_{n \rightarrow \infty} 
{\pi e^2\over{\mu c^2}}\cdot
{f_{1s,np}\over{\Delta\tilde{\nu}}} = {\pi e^2\over{\mu c^2}}\cdot
2^7\exp(-4)\cdot {1\over{3Ry}}. 
\end{equation}

\noindent 
If we work out the cross section from first principles, we have 

\begin{equation}
\sigma_{\rm lim} = {8\pi^3\over 3}{\nu\over c}a_H^3\cdot
|<1,s|{\bf r}|E=0, p>|^2.
\end{equation}

The dimensionless matrix element for a freed electron with zero
energy has the absolute numerical value 2.165364.  
Both Equations (4) and (5)
agree for the cross section for the series limit given in Table 1,
as they must.

We omit details of a corresponding proof that we have carried
out for the Balmer series.  It is necessary to convert the
closed-form expressions for the relevant radial integrals 
(cf. Bethe and Salpeter, Eqs. 63.4) to $f$-values, and perform
an average for the $2s -np$, $2p - nd$, and $2p - ns$ series:

\begin{equation}
<\sigma_{2-n}> = {1\over 4}[\sigma_{2s-np}+3\sigma_{2p-nd}
+3\sigma_{2p-ns}].
\end{equation}

We thank Drs. D. J. Bord, L. L. Lohr, P. J. Mohr, and  W. L. Wiese 
for useful comments.   
\vspace{0.2 in}
\begin{center}
{\it References}
\end{center}
\vspace{0.2 in}
\hangpar (1) C. Stehl\'{e} \& S. Jacquemot, {\it A\&A}, {\bf 271},
348, 1993.

\hangpar (2) D. G. Hummer \& D. Mihalas, {\it ApJ}, {\bf 331},
794, 1988.

\hangpar (3) H. R. Griem, {\it Principles of Plasma
Spectroscopy}, (Cambridge: University Press), 1997.

\hangpar (4) http://physics.nist.gov/cuu/Constants/index.html.
See also P. J. Mohr \& B. N. Taylor, {\it Rev. Mod. Phys.},
{\bf 72}, No. 2, 2000.

\hangpar (5) H. R. Bethe \& E. E. Salpeter, {\it Quantum
Mechanics of One- and Two-Electron Atoms}, (New York: Academic
Press), 1957.

\hangpar (6) L. C. Green, P. P. Rush \& C. C. Chandler, {\it ApJS},
{\bf 3}, 37, 1957.

\hangpar (7) W. J. Karzas \& R. Latter, {\it ApJS}, {\bf 6}, 167,
1961.

\hangpar (8) W. L. Wiese, D. E. Kelleher \& D. R. Paquette,
{\it Phys. Rev.}, {\bf A6}, 1132, 1972.

\hangpar (9) M. J. Seaton, {\it J. Phys. B}, {\bf 23}, 3255, 1990.

\end{document}